\documentclass[12pt]{article}

\usepackage[utf8]{inputenc}

\usepackage{amsmath}
\usepackage{amsfonts}
\usepackage{color}
\usepackage{graphicx}

\newtheorem{thm}{Theorem}[section]

\newtheorem{prop}[thm]{Proposition}

\newcommand{\R}{\mathbb R}
\newcommand{\Pb}{\mathbb P}
\newcommand{\Esp}{\mathbb E}

\title{Finite state space non parametric Hidden Markov Models are in general identifiable}
\author{E. Gassiat, A. Cleynen and S. Robin}
\date{\today}

\begin{document}

%%%%%%%%%%%%%%%%%%%%%%%%%%%%%%%%%%%%%%%%%%%%%%%%%%%%%%%%%%%%%%%%%%%%%%%%%%%%%%%%%%
\maketitle

\begin{abstract}
In this paper, we prove that finite state space non parametric hidden Markov models are identifiable as soon as the transition matrix of the latent Markov chain has full rank and the emission probability distributions are linearly independent. We then propose several non parametric likelihood based estimation methods, which we apply to models used in applications. We finally show on examples that the use of non parametric modeling and estimation may improve the classification  performances.
\end{abstract}

\section{Introduction}
\label{sec:intro}

Finite mixtures are widely used in applications to model data coming from different populations.
Let $X$ be the latent random variable whose value is the label of the population the observation comes from, and let $Y$ be the observed random variable.
With finitely many populations, $X$ takes values in $\{1,\ldots,k\}$ for some fixed integer $k$, and conditionally to $X=j$, $Y$ has distribution $\mu_{j}$.
Here, $\mu_{1},\ldots,\mu_{k}$ are probability distributions on the observation space ${\cal Y}$ endowed with its Borel sigma-field and are called emission distributions. Assume that we are given $n$ observations $Y_{1},\ldots,Y_{n}$ with the same distribution as $Y$, that is with distribution
\begin{equation}
\label{mel}
\sum_{j=1}^{k}\pi_{j} \mu_{j}
\end{equation}
where $\pi_{j}=\Pb(X=j)$, $j=1,\ldots,k$. If the latent variables $X_{1},\ldots,X_{n}$ are i.i.d., then so are the observed variables $Y_{1},\ldots,Y_{n}$. If the latent variables are not independent, then the observed variables are not either. The dependency structure of the observed variables is given by that of the latent variables.

%Independent variables were mainly used in the early XXXX. 
To be able to infer about the population structures, that is about the emission distributions, one usually states parametric models, saying that the emission distributions belong to some set parametrized by finitely many parameters (for instance Poisson distributions, or Gaussian distributions). Indeed, one may not recover the individual emission distributions from a convex combination of them without further information.
Since independence is often a crude approximation of the joint behavior of observed variables, mixture models were generalized to hidden Markov models to be used for clustering purposes. In hidden Markov models (shortened as HMMs in the paper), the latent variables form a Markov chain. Efficient algorithms allow to compute the likelihood and to build practical inference methods, see \cite{cappe:ryden:2005} for a recent state of the art in HMMs.

But parametric modeling of emission distributions may lead to poor results, in particular for clustering purposes.
Recent interest in using nonparametric HMMs appeared in applications, see for instance \cite{CC00} for voice activity detection, \cite{LWM03} for climate state identification, \cite{Lef03} for automatic speech recognition, \cite{SC09} for facial expression recognition, \cite{VBM13} for methylation comparison of proteins. These papers propose algorithms to get non parametric estimators and perform classification, but none of them gives theoretical results to support the methods. As noted before, theoretical results in the independent context may only be obtained under further assumptions on the emission distributions. But it has been proved recently by one of the authors (\cite{gassiat:rousseau:13}) that, for translations mixtures, that is when the emission distributions are all translated from an unknown one, identifiability holds without any assumption on the translated distribution provided that the latent variables are indeed not independent.

The aim of this paper is to prove that for HMMs, this result may be generalized to any mixtures with Markov regime. As a consequence, consistent estimators of the distribution of the latent variables and of the emission distributions may be built, leading to non parametric classification procedures.
In some sense, the message of our paper may be summarized as: {\it non independence of the observed variables helps if one wants to identify the population structure of the data and to cluster the observations}.

In section \ref{sec:ident} we state and prove our main result: non parametric HMMs may be fully identified provided the transition matrix of the hidden Markov chain has full rank, and the emission distributions are linearly independent, see Theorem \ref{thm:identifiability}. 
We then propose various estimation procedures, that should be consistent, thanks to identifiability.
In section \ref{sec:models} we show how our identifiability result applies to models used in applications.
Finally, in section \ref{sec:simuls} we present a simulation study mimicking RNA-Seq data and an application to transcriptomic tiling array data.

\section{The identifiability result and consequences} \label{sec:ident}

\subsection{Main theorem}
Let $(X_{i})_{i\geq 1}$ be a stationary Markov chain on $\{1,\ldots,k\}$. 
Let $(Y_{i})_{i\geq 1}$ be a real valued HMM, that is, a sequence of real valued random variables such that, conditionally to $(X_{i})_{i\geq 1}$, the $Y_{i}$'s are independent, and their distribution depends only on the current $X_{i}$. 
If $Q$ is the transition matrix of the Markov chain, and $M=(\mu_{1},\ldots,\mu_{k})$ are $k$ probability distributions on $\R$, we denote by $\Pb_{Q,M}$ the distribution of $(Y_{i})_{i\geq 1}$, where $(X_{i})_{i\geq 1}$ has transition $Q$ and $\mu_{i}$ is the distribution of $Y_{1}$ conditionally to $X_{1}=i$, $i=1,\ldots,k$. We call $\mu_{1},\ldots,\mu_{k}$ the emission distributions. Notice that in case the Markov chain is irreducible, there exists a unique stationary distribution and $\Pb_{Q,M}$ is well defined, while in the case where the Markov chain is not irreducible, there might exist several stationary distributions so that the distribution of $X_{1}$ has to be specified. 

\begin{thm} \label{thm:identifiability}
Assume $k$ is known, that the probability measures  $\mu_{1},\ldots,\mu_{k}$ are linearly independent, and that $Q$ has full rank. Then the parameters $Q$ and $M$ are identifiable from $\Pb_{Q,M}^{(3)}$, that is from the distribution of $3$ consecutive observations $Y_{1},Y_{2},Y_{3}$, up to label swapping of the hidden states.
\end{thm}

Let us prove Theorem \ref{thm:identifiability}.
We have to prove that, if $\tilde{Q}$ is a $k\times k$ transition matrix, and if $\tilde{M}=(\tilde{\mu}_{1},\ldots,\tilde{\mu}_{k})$ are $k$ probability distributions on $\R$,  if $\Pb_{\tilde{Q},\tilde{M}}^{(3)}=\Pb_{Q,M}^{(3)}$, then there exists a permutation $\sigma$ of the set $\{1,\ldots,k\}$ such that, for all $i,j=1,\ldots,k$, $\tilde{Q}_{i,j}=Q_{\sigma(i),\sigma(j)}$ and $\tilde{\mu}_{i}=\mu_{\sigma(i)}$.\\
%This follows easily from
%Theorem 8 of
%\cite{AllMatRho09} and the fact that, 
Now, since  $(X_{i})_{i\geq 1}$ is a Markov chain, conditionally to $X_{2}$, $X_{1}$ and $X_{3}$ are independent variables.  Then, using this fact, the distribution of $(Y_{1},Y_{2},Y_{3})$ under $\Pb_{Q,M}$ may be written as
$$
\Pb_{Q,M}^{(3)}=\sum_{i=1}^{k} \left(\sum_{j=1}^{k}\pi_{j}Q_{j,i} \mu_{j}\right)\otimes \mu_{i} \otimes \left(\sum_{j=1}^{k}Q_{i,j}\mu_{j}\right),
$$
where $(\pi_{1},\ldots,\pi_{k})$ is a stationary distribution of $Q$ which is the distribution of $X_{1}$.
Similarly,
$$
\Pb_{\tilde{Q},\tilde{M}}^{(3)}=\sum_{i=1}^{k} \left(\sum_{j=1}^{k}\tilde{\pi}_{j}\tilde{Q}_{j,i} \tilde{\mu}_{j}\right)\otimes \tilde{\mu}_{i} \otimes \left(\sum_{j=1}^{k}\tilde{Q}_{i,j}\tilde{\mu}_{j}\right),
$$
where $(\tilde{\pi}_{1},\ldots,\tilde{\pi}_{k})$ is a stationary distribution of $\tilde{Q}$  which is the distribution of $X_{1}$ under $\Pb_{\tilde{Q},\tilde{M}}$.
%
% $\nu^{1}_{i}$ is the distribution of $Y_{1}$ conditionally to $X_{2}=i$ and $\nu^{3}_{i}$ is the distribution of $Y_{3}$ conditionally to $X_{2}=i$. Notice that since $Q$ has full rank, for all $i=1,\ldots,k$, $\pi_{i}>0$. One has
%$$
%\nu^{1}_{i}=\sum_{j=1}^{k}\frac{\pi_{j}Q_{ji}}{\pi_{i}} \mu_{j},
%$$
%and using the fact that $Q$ has full rank and the  probability measures  $\mu_{1},\ldots,\mu_{k}$ are linearly independent, we easily get that  the probability measures  $\nu^{1}_{1},\ldots,\nu^{1}_{k}$ are linearly independent. Also,
%$$
%\nu^{3}_{i}=\sum_{j=1}^{k}Q_{ij}\mu_{j},
%$$
Since $Q$ has full rank and the  probability measures  $\mu_{1},\ldots,\mu_{k}$ are linearly independent, the measures  $\left(\sum_{j=1}^{k}\pi_{j}Q_{j,i} \mu_{j}\right)$, $i=1,\ldots,k$ are linearly independent, and the probability measures $\left(\sum_{j=1}^{k}Q_{i,j}\mu_{j}\right)$,  $i=1,\ldots,k$ are also linearly independent.
Thus, applying Theorem 8 of
\cite{AllMatRho09} we get that there exists a permutation $\sigma$ of the set $\{1,\ldots,k\}$ such that, for all $i=1,\ldots,k$ :
$$
\sum_{j=1}^{k}\tilde{\pi}_{j}\tilde{Q}_{j,i} \tilde{\mu}_{j}=\sum_{j=1}^{k}\pi_{j}Q_{j,\sigma(i)} \mu_{j},\;
\tilde{\mu}_{i}=\mu_{\sigma(i)},\;
\sum_{j=1}^{k}\tilde{Q}_{i,j} \tilde{\mu}_{j}=\sum_{j=1}^{k}Q_{\sigma(i),j} \mu_{j}.
$$
This gives easily,  for all $i=1,\ldots,k$,
$$
\sum_{j=1}^{k}\tilde{\pi}_{j}\tilde{Q}_{j,i} \mu_{\sigma(j)}=\sum_{j=1}^{k}\pi_{\sigma(j)}Q_{\sigma(j),\sigma(i)} \mu_{\sigma(j)},\;
\sum_{j=1}^{k}\tilde{Q}_{i,j} \mu_{\sigma(j)}=\sum_{j=1}^{k}Q_{\sigma(i),\sigma(j)} \mu_{\sigma(j)}.
$$
Using now the linear independence of $\mu_{1},\ldots,\mu_{k}$ we get that  for all $i,j=1,\ldots,k$,
$$
\tilde{Q}_{j,i}=Q_{\sigma(j),\sigma(i)} ,\;\tilde{\pi}_{j}\tilde{Q}_{j,i} =\pi_{\sigma(j)}Q_{\sigma(j),\sigma(i)} 
$$
and the theorem is proved.

\subsection{Non parametric estimation}
Identifiability is the building stone for estimation procedures to lead to consistent estimators. We may now propose several estimation procedures. Let us set the ideas for likelihood based procedures, for which the popular EM algorithm may be used to compute the estimators, as we recall in Section \ref{subsec:EM}. 
Assume that the emission distributions are dominated by a measure $\nu$ on $\cal Y$. Let $\theta=(Q,f_{1},\ldots,f_{k})$, $f_{j}$ being the density of $\mu_{j}$ with respect to the dominating measure. Then $(Y_{1},\ldots,Y_{n})$ has a density $p_{n,\theta}$ with respect to $\nu^{\otimes n}$. Denote
$
\ell_{n}\left(\theta\right)=\log p_{n,\theta} \left( Y_{1},\ldots,Y_{n}\right)
$ the log-likelihood, and $
\tilde{\ell}_{n}\left(\theta\right)=\sum_{i=1}^{n-2}\log p_{3,\theta} \left( Y_{i},Y_{i+1},Y_{i+2}\right)
$
 the pseudo log-likelihood. Likelihood (or pseudo-likelihood) based non parametric estimation usually involves a penalty, which might be chosen as a regularization term (as studied in \cite{vdG:00} mainly for independent observations)
or as a model selection term (see  \cite{massart:03}). More precisely:
\begin{itemize}
\item
Let $I(f)$ be some functional on the density $f$. For instance, if $\cal Y$ is the set of non negative integers, one may take $I(f)=\sum_{j\geq 0} j^{\alpha} f(j)$  for some $\alpha >0$; if $\cal Y$ is the set of real numbers, one may take $I(f)=\int_{-\infty}^{+\infty}[f^{(\alpha)}(u)]^{2} du$, where $f^{(\alpha)}$ is the $\alpha$-th derivative of $f$.
Then the estimator may be chosen
as a maximizer of 
\begin{equation} \label{eq:penloglik}
\ell_{n}\left(\theta\right)-\lambda_{n}[I(f_{1})+\ldots +I(f_{k})],
\end{equation}
or of $\tilde{\ell}_{n}\left(\theta\right)-\lambda_{n}[I(f_{1})+\ldots +I(f_{k})]$ for some well chosen positive sequence $(\lambda_{n})_{n\geq 1}$. In Section \ref{subsec:discrete} we provide an application of this estimator which we further illustrate in Section \ref{subsec:RNAseq}. 
\item
If we consider for $\theta$ a sequence of models $({\Theta}_{m})_{m\in {\cal M}}$ where ${\Theta}_{m}$ is the set of possible values for $\theta$ for constraint $m$, one may choose the estimator of $m$ as a maximizer over ${\cal M}$ of $\ell_{n}\left(\widehat{\theta}_{m}\right)-\text{pen}(n,m)$ (or of $\tilde{\ell}_{n}\left(\widehat{\theta}_{m}\right)-\text{pen}(n,m)$), where $\text{pen}(n,m)$ is some penalty term. Here, $\widehat{\theta}_{m}$ is the maximum likelihood estimator (or the maximum pseudo-likelihood estimator) in model ${\Theta}_{m}$ for each $m\in {\cal M}$. In Section \ref{subsec:mix} we consider for models ${\Theta}_{m}$ the set of the emission densities which can be modeled as mixture distributions with $m$ components.
%If we consider a sequence of models $({\cal M}_{m})_{m\in M}$ for $\theta$, one may choose the estimator as a maximizer of $\ell_{n}\left(\widehat{\theta}_{m}\right)-\text{pen}(n,m)$ (or of $\tilde{\ell}_{n}\left(\widehat{\theta}_{m}\right)-\text{pen}(n,m)$), where $\text{pen}(n,m)$ is some penalty term. Here, $\widehat{\theta}_{m}$ is the maximum likelihood estimator (or the maximum pseudo-likelihood estimator) in model ${\cal M}_{m}$ for each $m\in M$. In Section \ref{subsec:mix} we consider models ${\cal M}_{m}$ in which the emission densities or modeled as mixture distributions with $m$ components.
\end{itemize}

We may also consider usual non parametric estimators for emission densities. For instance, in Section \ref{subsec:ker} we consider kernel based estimators computed via maximum likelihood, which we illustrate in Section \ref{subsec:tiling}.

%To get theoretical results, one 
%has to use deviation inequalities, which are known for additive functionals of Markov chains together with the usual techniques to obtain rates of convergence for
%the square Hellinger distance between $ p_{3,\theta}$ and $ p_{3,\widehat{\theta}}$.

\section{Application to some specific models} \label{sec:models}

In this section we present and discuss a series of hidden Markov models that can be proved to be identifiable thanks to the results above.

%%%%%%%%%%%%%%%%%%%%%%%%%%%%%%%%%%%%%%%%%%%%%%%%%%%%%%%%%%%%%%%%%%%%%%%%%%%%%%%%%%
\subsection{Reminder on the inference of hidden Markov models}
\label{subsec:EM}

A huge variety of techniques have been proposed for the inference of hidden Markov models (see e.g. \cite{cappe:ryden:2005}). The most widely used is probably the E-M algorithm proposed by \cite{DLR77}, which can be adapted to several illustrations given below. We recall that this algorithm alternates an expectation (E) step with a maximization (M) step until convergence. At iteration $h+1$, the (M) step retrieves estimates $Q^{h+1}$ and $M^{h+1}$ via the maximization of the conditional expectation
\begin{eqnarray} \label{eq:FQM}
F^h(Q, M) & = & 
\Esp_{Q^h, M^h}\left[\log p_{n,(Q, M)}((Y_i)_{1\leq i \leq n}, (X_i)_{1\leq i \leq n}) | (Y_i)_{1\leq i \leq n} \right] \nonumber \\
& = & 
\Esp_{Q^h, M^h}\left[\log p_{n,(Q,M)}((X_i)_{1\leq i \leq n}) | (Y_i)_{1\leq i \leq n} \right] \nonumber \\
& + &
\Esp_{Q^h, M^h}\left[\log p_{n,(Q,M)}((Y_i)_{1\leq i \leq n} | (X_i)_{1\leq i \leq n}) | (Y_i)_{1\leq i \leq n} \right] 
\end{eqnarray}
w.r.t. $Q$ and $M$. This expectation involves the current estimates of the conditional probabilities: $\tau_{ij}^h := \Pb_{Q^h, M^h}(X_i=j | (Y_i)_{1\leq i \leq n})$ and $\Pb_{Q^h, M^h}(X_i=j, X_{i+1}=j' | (Y_i)_{1\leq i \leq n})$. These conditional probabilities are updated at the next (E) step, using the forward-backward recursion, which takes the current parameter estimates $Q^{h+1}$ and $M^{h+1}$ as inputs. In the sequel, we focus on the estimation of $M$, the rest of the calculations being standard.

%%%%%%%%%%%%%%%%%%%%%%%%%%%%%%%%%%%%%%%%%%%%%%%%%%%%%%%%%%%%%%%%%%%%%%%%%%%%%%%%%%
\subsection{Non-parametric discrete distributions}\label{subsec:discrete}

We consider a hidden Markov model with discrete observations $(Y_i)_{i\geq 1}$ with fully non parametric emission distributions $\mu_j$ (denoting $f_j(y) = \Pb(Y_i = y | X_i = j) $).
% , each with respective support ${\mathcal S}_j \subseteq {\mathbb Z}$. \SR{Faut-il parler de l'estimation du support ou simplement supposer que $\mathcal S_j = \mathbb Z$ ?} 
Theorem \ref{thm:identifiability} ensures that, provided that the distributions $\mu_j$ are all linearly independent, the corresponding HMM is identifiable.

%\SR{Theorem 1 deals with continuous distributions.}

\paragraph{Inference.} The maximum likelihood inference of this model can be achieved via EM, the M step resulting in
$$
f_j^h(y) = S^h_j(y) / N_j^h
$$
where $S^h_j(y) = \sum_i \tau_{ij}^h {\mathbb I}(Y_i=y)$ and $N_j^h = \sum_i \tau_{ij}^h$.

\paragraph{Regularization.} The EM algorithm can be adapted to the maximization of a penalized likelihood such as \eqref{eq:penloglik}. Indeed the regularization only affects the (M) step (see \cite{li2005regularized}). Taking $I(f) = \sum_y m(y) f(y)$ (e.g. $m(y) = y^\alpha$), the estimate of $f_j$ satisfies
$$
f_j^h(y) = S^h_j(y) \left/ \left(\lambda_n m(y) + c^h_j \right) \right. 
$$
where the constant $c^h_j$ ensures that $\sum_y f^h_j(y) = 1$. Note that this estimate is not explicit but, as $\sum_y f^h_j(y)$ is a monotonous decreasing function of $c^h_j$, this constant can be efficiently determined using any standard algorithm, such as dichotomy.

\paragraph{RNA-Seq data.} In the past few years, next generation sequencing (NGS) technologies have become the state-of-the-art tool for a series of applications in molecular biology such as transcriptome analysis, giving raise to RNA-Seq. Briefly speaking, NGS provide reads that can be aligned along a reference genome, so that a count is associated with each nucleotide. The resulting RNA-Seq count is supposed to reveal the level of transcription of the corresponding nucleotide. HMMs have been proposed (\cite{du2006supervised,zhai2010power}) to determine transcribed regions based on RNA-Seq. The choice of the emission distribution is one of the main issue of such modeling. Poisson distributions display a poor fit to the observed data and the negative binomial has emerged as the the consensus distribution. However, no theoretical justification for such a model exists. Furthermore, the inference of negative binomial models raises several problems, especially for the over-dispersion parameter. The simulation study we 
perform in  Section \ref{sec:simuls} shows that fully non
parametric emission distributions can be used and improve the classification performances.
% \SR{A répartir avec l'intro de l'application ?}

%%%%%%%%%%%%%%%%%%%%%%%%%%%%%%%%%%%%%%%%%%%%%%%%%%%%%%%%%%%%%%%%%%%%%%%%%%%%%%%%%%
\subsection{Mixtures as emission distributions}
\label{subsec:mix}

Latent variable models with parametric emission distributions often poorly fit the observed data due to the choice of the emission distribution. In the recent years, big efforts have been made to consider more flexible parametric emission distributions (see e.g. \cite{LLY07}). Mixture distribution have recently been proposed to improve flexibility (see \cite{baudry2010combining}). The model is the following: consider a set of $m$ parametric distributions $\phi_\ell$ ($\ell = 1 \dots m$) and a $k\times m$ ($m \geq k$) matrix of proportions $\psi = [\psi_{i\ell}]$ such that, for all $j = 1 \dots k$, $\sum_\ell \psi_{j\ell} = 1$. The emission distribution $\mu_j$ is defined as
\begin{equation} \label{eq:mixtemission}
\mu_j = \sum_\ell \psi_{j\ell} \phi_\ell.
\end{equation}
A simple mixture model (i.e. when the hidden variable $X_i$ are iid) with such emission distribution is not identifiable (see \cite{baudry2010combining}). However, its hidden Markov model counterpart is identifiable, under the conditions stated in the following proposition.

\begin{prop}
\label{prop:mix}
 If the distributions $\phi_\ell$ are linearly independent and if the matrix $\psi$ has rank $k$, then the HMM with emission distribution $\mu_j$ defined in \eqref{eq:mixtemission} is identifiable as soon as $Q$ has also full rank.
\end{prop}

{\bf Proof.} As the distributions $\phi_\ell$ are linearly independent, it suffices that the rows of $\psi$ are linearly independent to ensure that so do the distributions $\mu_j$. Identifiability then results from Theorem \ref{thm:identifiability}. $\square$

%\paragraph{LOH}

\paragraph{Inference.} The maximum likelihood inference of such a model has been studied in \cite{VBM13}, although identifiability issues are not theoretically addressed therein. The EM algorithm can be adapted to this model, considering a second hidden sequence of variables $Z_1, \dots, Z_n$ that are independent conditional on the $(X_i)$ each with multinomial distribution:
$$
(Z_i | X_i=j) \sim \mathcal M(1; \psi_j)
$$
where $\psi_j$ stands for the $j$th row of $\psi$. Note that the sequence  $Z_1, \dots, Z_n$ is itself a Markov chain, so the conditional probability $\xi_{i\ell}^h := \Pb(Z_i=\ell | (Y_i)_{i \geq 1})$ can be computed via the forward-backward recursion during the (E) step. See \cite{VBM13}.

\paragraph{Mixture of exponential family distributions.} 
In such modeling, the distributions $\phi_\ell$ are often chosen within the exponential family, that is
$$
\phi_\ell(y) = \exp[\theta'_\ell t(y) - a(y) - b(\theta_\ell)]
$$
where $t(y)$ stands for the vector of sufficient statistics, $\theta_\ell$ for the vector of canonical parameters and $a$ and $b$ for the normalizing functions. Standard properties of maximum likelihood estimates in the exponential family yield that the estimates of $\theta^h_\ell$ resulting from the M step must satisfy
$$
b'(\theta^h_\ell) = T^h_\ell/ N_\ell^h
$$
where $T^h_\ell = \sum_i \xi_{i\ell}^h t(Y_i)$ and $N_j^h = \sum_i \xi_{i\ell}^h$. Explicit estimates result from this identity for a series of distribution such as multivariate Gaussian, Poisson, or Binomial. Indeed, Gaussian, Poisson and Binomial ${\cal B}(N,p)$ for $N\geq 2m-1$ distributions are linearly independent, as shown in \cite{TSM85}.

%%%%%%%%%%%%%%%%%%%%%%%%%%%%%%%%%%%%%%%%%%%%%%%%%%%%%%%%%%%%%%%%%%%%%%%%%%%%%%%%%%
\paragraph{Convex emission distribution}

Discrete convex distributions are proved in
\cite{DHKR13} to be mixtures of triangular discrete distributions. It may be proved, in the same way as in Theorem 8 of \cite{DHKR13} that those triangular discrete distributions are in fact linearly independent so that one may use Proposition \ref{prop:mix}.

\paragraph{Zero-inflated distributions}

Zero-inflated distributions are mixtures of a Dirac delta distribution $\delta_0$ and a distribution $\phi_j$, which is typically chosen from but not limited to the exponential family, so that the emission distribution $\mu_j$ can be defined as
$$
\mu_j = q_j \delta_0 + (1-q_j) \phi_j.
$$

This model can be expressed as a particular case from that of Equation (\ref{eq:mixtemission}) for which $m=k+1$ and $\phi_{k+1}=\delta_0$. The matrix $\psi$ is then sparse, with last column $q=(q_1,\dots,q_k)$ and main diagonal $1-q$. This ensures that provided at most one $q_j$ is equal to one, $\psi$ has full rank. It thus suffices that the $\phi_j$ are linearly independent to allow the use of Proposition \ref{prop:mix}, and give support to a vast literature (see \cite{desantis2011hidden,olteanu2012hidden} for examples of usage of zero-inflated Poisson HMMs to model over-dispersed count datasets).

\paragraph{Non parametric density modeling via mixtures}
Mixtures, in particular Gaussian mixtures, may be used for a model selection approach for the non parametric estimation of probability densities, see \cite{MB11}. See also
\cite{gassiat:rousseau:13}  in the HMM context.

%%%%%%%%%%%%%%%%%%%%%%%%%%%%%%%%%%%%%%%%%%%%%%%%%%%%%%%%%%%%%%%%%%%%%%%%%%%%%%%%%%
\subsection{Kernel density estimation}
\label{subsec:ker}

Two major classes of nonparametric density estimators for continuous variables are proposed in the literature in an attempt at capturing the specific shapes of the data where parametric approaches fail: kernel estimates, of which the histogram approach presented in Section \ref{subsec:discrete} is a special case, and wavelet-based techniques. We refer to \cite{donoho1996density} for a complete description of wavelet-estimates properties, or \cite{CC00} for an example of their use in non-parametric HMMs. 

We will focus on kernel-based estimates for the emission densities and for a given bandwith $w$, we will write $f_j(y)$ of the form
$$
f_j(y) = \frac1{w} \sum_u p_{uj} R\left(\frac{y - y_u}w\right)
$$
where $R$ is some symmetric kernel function satisfying $\int R= 1$ and where the $p_{uj}$ are weights such that, for all $u$, $\sum_u p_{uj} = 1$. A similar estimate was proposed by \cite{HallZhou03}. We denote $P = (p_{uj})$ the set of all weights. In this setting, for a given $w$, the estimation of $(f_{1},\ldots,f_{k})$ amounts to the estimation of $P$. 

\paragraph{Maximum likelihood.} An EM algorithm can be used to get maximum likelihood estimates of $Q$ and $P$. We define 
$$
G^h(P) = \Esp_{Q^h, M^h}\left[\log p_{n,(Q,M)}((Y_i)_{1\leq i \leq n} | (X_i)_{1\leq i \leq n}) | (Y_i)_{1\leq i \leq n} \right], 
$$
which corresponds to the last term of \eqref{eq:FQM} and is the only term  to depend on $P$. As for the estimation of $P$, the (M) step aims at maximizing this function that can be rewritten as
\begin{eqnarray} \label{eq:FQP}
G^h(P) & = & \sum_{i, j} \tau^h_{ij} \log \left(\frac1{w} \sum_u p_{uj} R_{iu} \right) \nonumber \\
%  & = & \sum_{i, j} \tau^h_{ij} \log \left(\sum_u p_{uj} k_{iu}\right) - n \log (w) \nonumber \\
  & = &  \sum_{i, u, j} \tau^h_{ij} \gamma_{iuj} \log \left(p_{uj} R_{iu} \right) - \sum_{i, u, j} \tau^h_{ij} \gamma_{iuj} \log \gamma_{iuj} - n \log (w)
\end{eqnarray}
where $R_{iu} = R((Y_i - Y_u)/w)$ and $\gamma_{iuj} = p_{uj} R_{iu} / \sum_v p_{vj} R_{iv}$.

\begin{prop} \label{prop:puj}
 The following recursion provides a sequence of increasing value of $G^h$:
 $$
 \gamma^{\ell}_{iuj} = p^\ell_{uj} R_{iu} / \sum_v p^\ell_{vj} R_{iv}, 
 \qquad
 p^{\ell+1}_{uj} = \sum_i \tau^h_{ij} \gamma^{\ell}_{iuj} \left/ \sum_{i, v} \tau^h_{ij} \gamma^{\ell}_{ivj}\right.
 $$
 satisfies $G^h(P^{\ell+1}) \geq G^h(P^{\ell})$.
\end{prop}

The proof of this proposition  is postponed to Appendix \ref{sec:proofs}. Iterating this recursion therefore improves the objective function $F^h(Q, M)$ --~even convergence is not reached~--, which results in a Generalized EM algorithm (GEM: \cite {DLR77}).\newline

% In practice, fitting this \textit{full} HMM model is very costly in time as at each $M$ step, a recursion is needed for the update of $p^\ell_{iu}$ until $G^h(P^{\ell})$ converges. 
% In the literature, different approaches have been proposed to accelerate the algorithm. One possibility is to reduce the number of iterations to a fixed (and low) number even if convergence is not reached, as required by the GEM algorithm. 
% Proposition \ref{prop:puj} ensures that $G^h(P^{\ell})$ increases at each step resulting in an increase of the likelihood at the $M$ step as required. 

Another common approach is to replace the terms $p_{uj}$ by the posterior probability that the $j^{th}$ individual belongs to class $\ell$.
This approximation is encountered in the non-parametric HMM literature both in kernel-based approaches (see for instance \cite{jin2006non}) and in wavelet-based approaches (see \cite{CC00}). 
However, even if this approximation is very intuitive (and much faster computationally), there is no theoretical guarantee that the EM-like algorithm increases the likelihood. 
In \cite{benaglia2009like} the authors show through simulation studies that it outperforms other approximation algorithm but fail to obtain descent properties. 
\cite{levine2011maximum} proposes a very similar algorithm, called Majorization-Minimization, which converges to a local maximum of a  smoothed likelihood.

\section{Simulation and application} \label{sec:simuls}

%%%%%%%%%%%%%%%%%%%%%%%%%%%%%%%%%%%%%%%%%%%%%%%%%%%%%%%%%%%%%%%%%%%%%%%%%%%%%%%%%%
\subsection{Simulation study}\label{subsec:RNAseq}
To study the improvement provided by the use of a non-parametric emission distributions, we designed a simulation study based on a typical application in genomics.

\paragraph{RNA-Seq data.}
Next generation sequencing (NGS) technologies allow to study gene expression all along the genome. NGS data consist of numbers of reads associated with each nucleotide. These read counts are function of the level of transcription of the considered nucleotide, so NGS allow to detect transcribed regions and to evaluate the level of transcription of each region. The state-of-the-art statistical methods are based on the negative binomial distribution.

\paragraph{Design.}
Based on the annotation of the yeast genome, we defined regions with four level of expression, from intronic (almost no signal) to highly expressed. We then used RNA-Seq data to define empirical count distributions for each of the four levels (so that $k=4$), which shall correspond to the hidden states. The data were simulated as follows: 14 regions were defined within a sequence of length $n = 1000$ and associated known states and the count at each position within this regions was sampled in the empirical distribution of the corresponding state. $S = 100$ synthetic datasets were sampled according to this scheme and we denote $y_i^s$ the observation from simulation $s$ ($s = 1\dots S$) at position $i$ ($i = 1 \dots n$).

\paragraph{Evaluation criteria.} For each simulation, three HMM models were fitted with, respectively, ($a$) negative binomial, ($b$) free non-parametric and ($c$) regularized emission distributions as defined in Section \ref{subsec:discrete}, taking
$$
I(f) = \sum_y y^2 f(y).
$$
For each model, we then inferred the hidden state $x_i^s$ according to both the maximum a posteriori (MAP) rule and the Viterbi most probable path. For each simulation, HMM and classification rule, we then computed the rand index between the inferred states $(\hat{x}_i)$ and the true one. We recall that the rand index is the proportion of concordant pairs of positions among the $n(n-1)/2$, where the pair $(i, i')$ is said concordant if either $x_i = x_{i'}$ and $\hat{x}_i = \hat{x}_{i'}$, or $x_i \neq x_{i'}$ and $\hat{x}_i \neq \hat{x}_{i'}$.
% $
% \mathbb I(x_i = x_{i'}) \mathbb I(\hat{x}_i = \hat{x}_{i'}) + \mathbb I(x_i \neq x_{i'}) \mathbb I(\hat{x}_i \neq \hat{x}_{i'}) = 1.
% $

%%%%%%%%%%%%%%%%%%%%%%%%%%%%%%%%%%%%%%%%%%%%%%%%%%%%%%%%%%%%%%%%%%%%%%%%%%%%%%%%%%
\subsection{Results}

\begin{figure}[h!]
  \begin{center}
    \begin{tabular}{ccc}
    \vspace{-.5cm}
    \includegraphics[width=.3\textwidth]{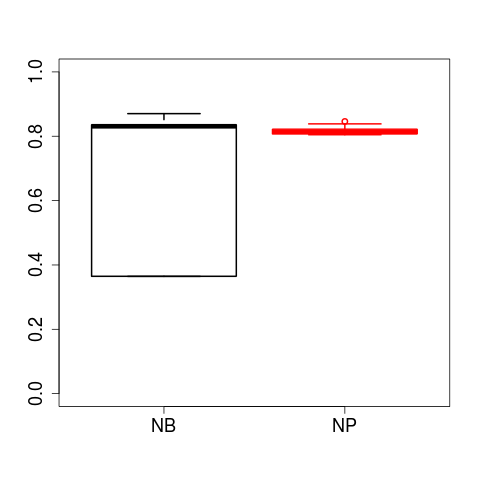} &
    \includegraphics[width=.3\textwidth]{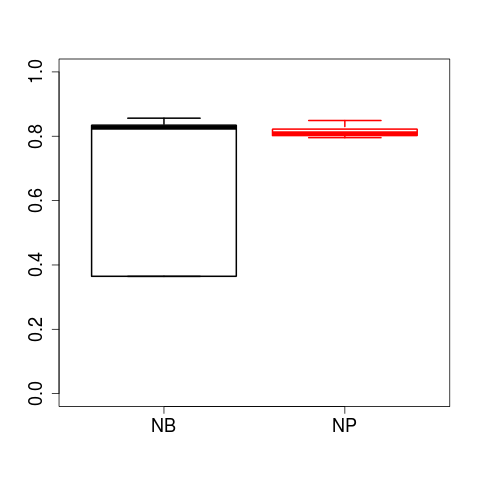} & 
    \includegraphics[width=.3\textwidth]{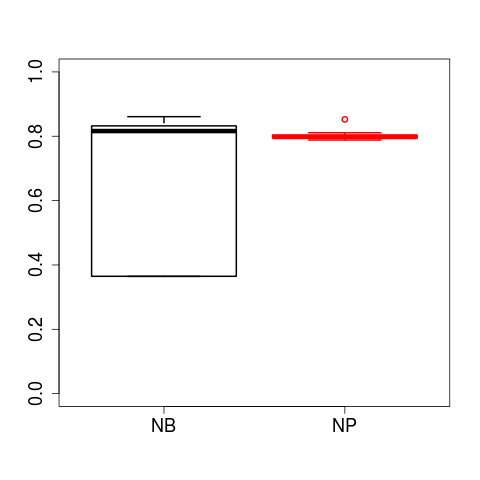} \\
    $k = 3$ & $k = 4$ & $k = 5$
    \end{tabular}
    \caption{Rand index for the two estimates: parametric negative binomial (NB: black) and non-parametric (NP: red).
    \label{fig:yeast-rand-NB-NP}}
  \end{center}
\end{figure}

MAP and Viterbi classifications achieved very similar performances so we only report the results for Viterbi. Figure \ref{fig:yeast-rand-NB-NP} displays the rand index for both the parametric (negative binomial) and non-parametric (with no regularization) estimates of the emission distribution, when considering three different values of $k$. We observe that, although the mean performances are similar with the two distributions, the parametric negative-binomial sometimes provides poor predictions. 

\begin{figure}[h!]
  \begin{center}
    \begin{tabular}{ccc}
    \vspace{-.5cm}
    \includegraphics[width=.3\textwidth]{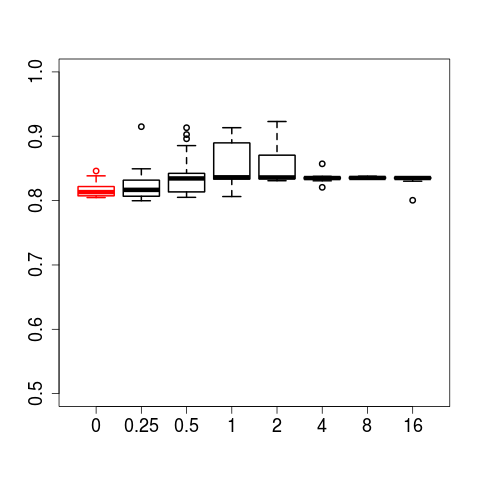} &
    \includegraphics[width=.3\textwidth]{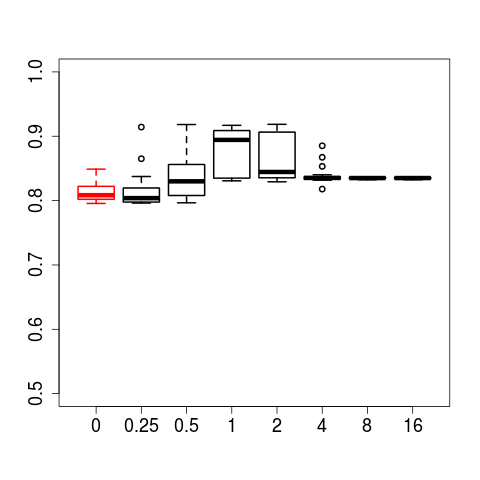} & 
    \includegraphics[width=.3\textwidth]{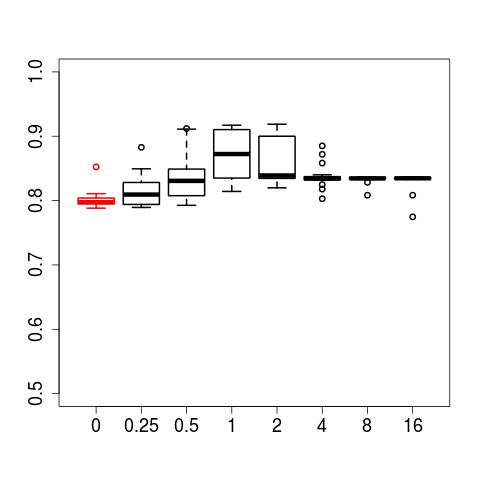} \\
    $k = 3$ & $k = 4$ & $k = 5$
    \end{tabular}
    \caption{Rand index as a function of the regularization parameter $\lambda$. $\lambda = 0$ (in red) corresponds to the non regularized estimate.
    \label{fig:yeast-rand-NP-rNP}}
  \end{center}
\end{figure}

We then studied the influence on regularization on the performances. We considered a set of values for $\lambda$, ranging from $0.25$ to $16$. Figure \ref{fig:yeast-rand-NP-rNP} shows that regularization can improve the results in a sensible manner. $\lambda = 1$ seems to work best in practice. We do not provide a systematic rule to choose the regularization parameter. Indeed, standard techniques such as cross-validation could be be considered, but would imply an important computational burden.

\begin{figure}
  \begin{center}
    \begin{tabular}{cc}
    \vspace{-.5cm}
    \includegraphics[width=.5\textwidth, height=.3\textheight]{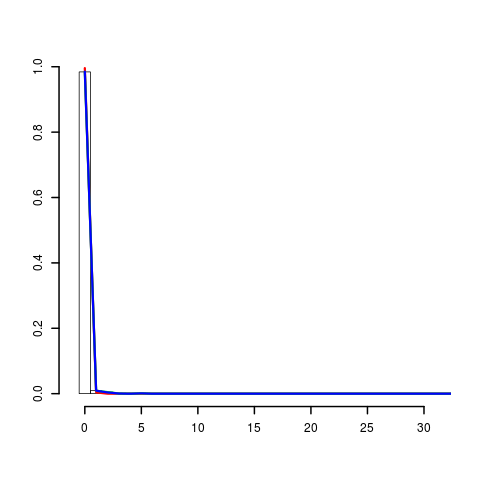} & 
    \includegraphics[width=.5\textwidth, height=.3\textheight]{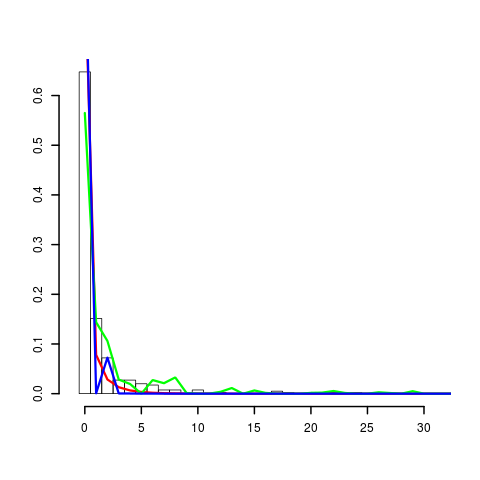} \\
    \includegraphics[width=.5\textwidth, height=.3\textheight]{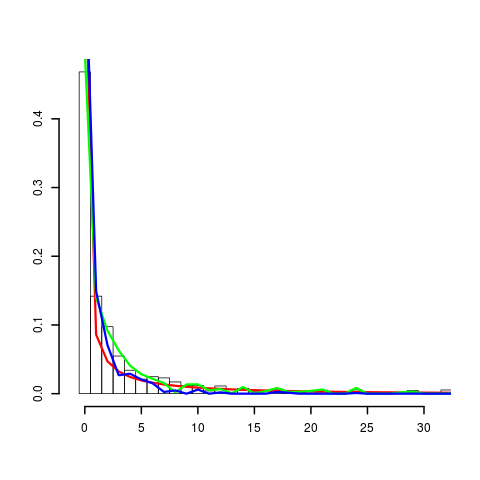} & 
    \includegraphics[width=.5\textwidth, height=.3\textheight]{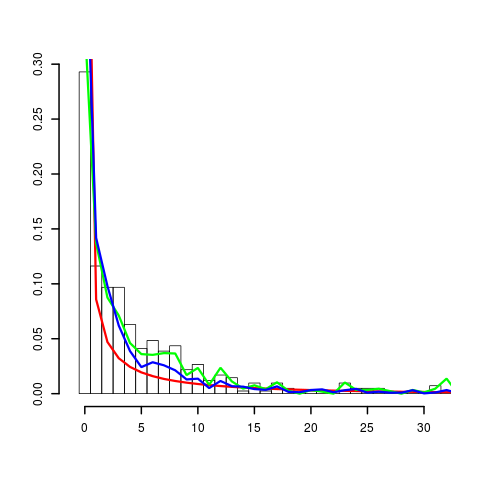} 
    \end{tabular}
    \caption{Fit of the estimated distributions with the three estimates: negative binomial (NB: red), non-parametric (NP: green) and regularized non-parametric ($\lambda = 1$, rNP: blue).
    \label{fig:yeast-fit}}
  \end{center}
\end{figure}

To illustrate the interest of the non-parametric estimate, we show in Figure \ref{fig:yeast-fit} the fits obtained with different estimates for a typical simulation. For the regularized version we used $\lambda = 1$ as suggested by the preceeding result. As expected, the unregularized estimate displays the best fit. 

\subsection{Application to transcriptomic tiling-array}\label{subsec:tiling}

\begin{figure}
  \begin{center}
    \begin{tabular}{ccc}
    \vspace{-.5cm}
    \includegraphics[width=.3\textwidth]{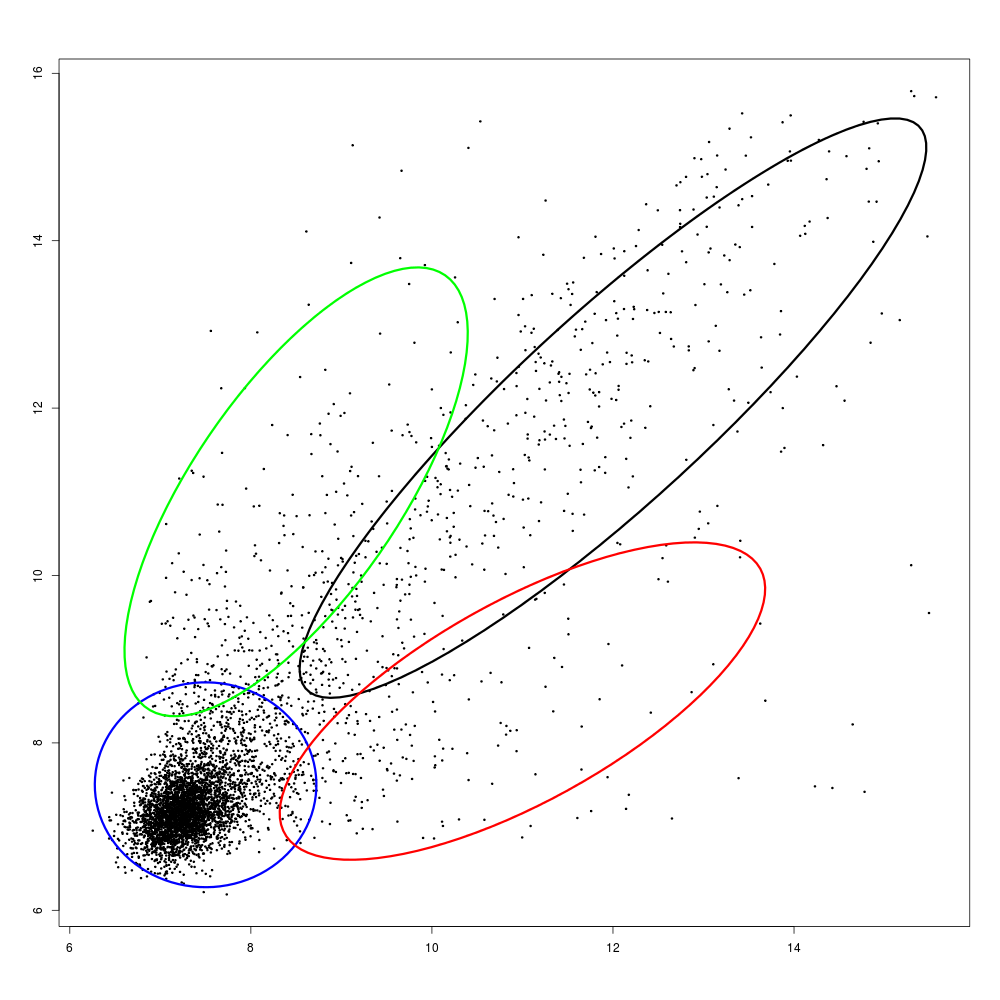} & 
    \includegraphics[width=.3\textwidth]{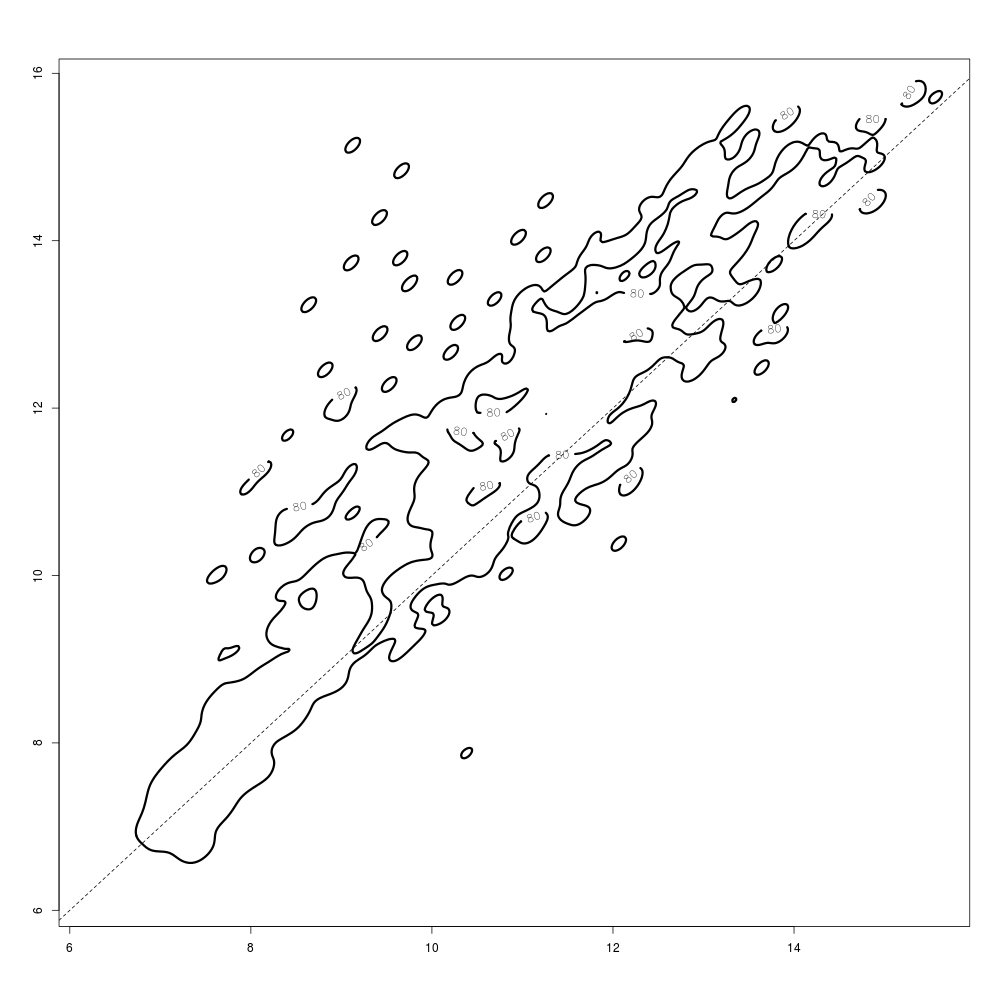} &
    \includegraphics[width=.3\textwidth]{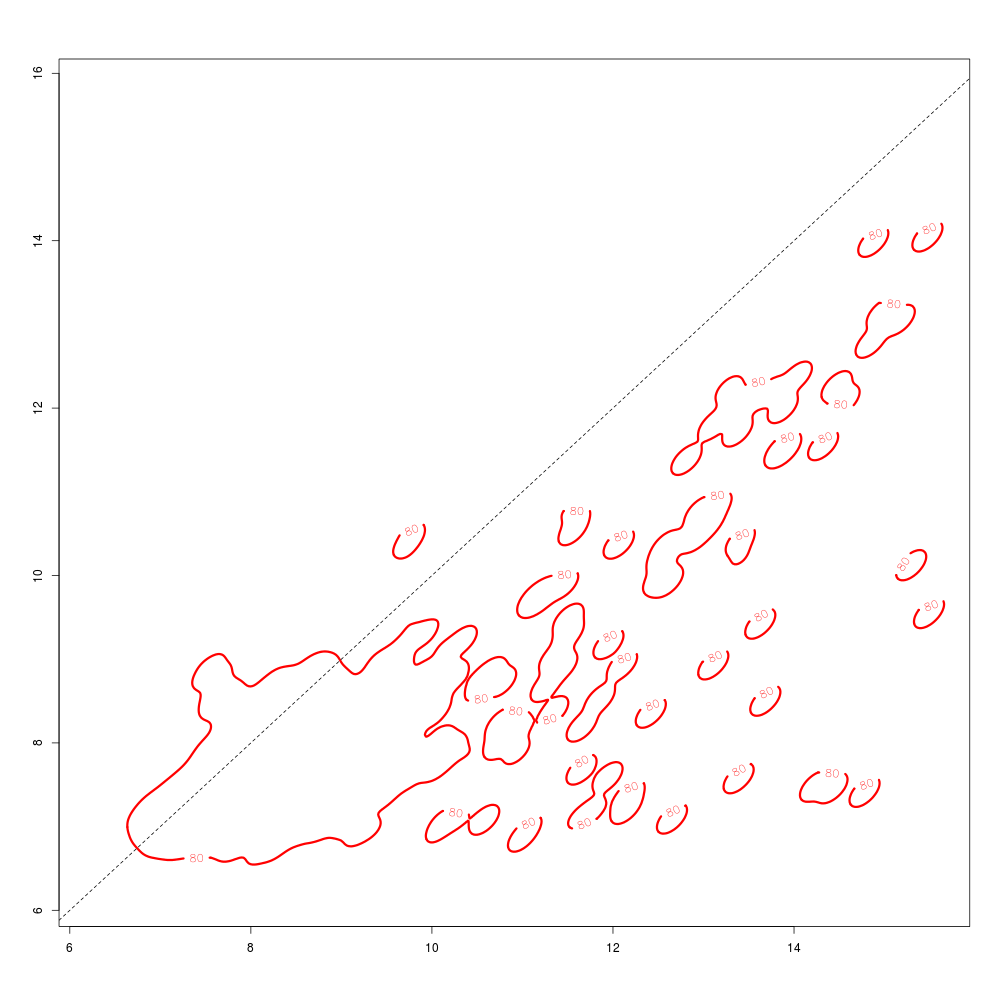} \\
    \includegraphics[width=.3\textwidth]{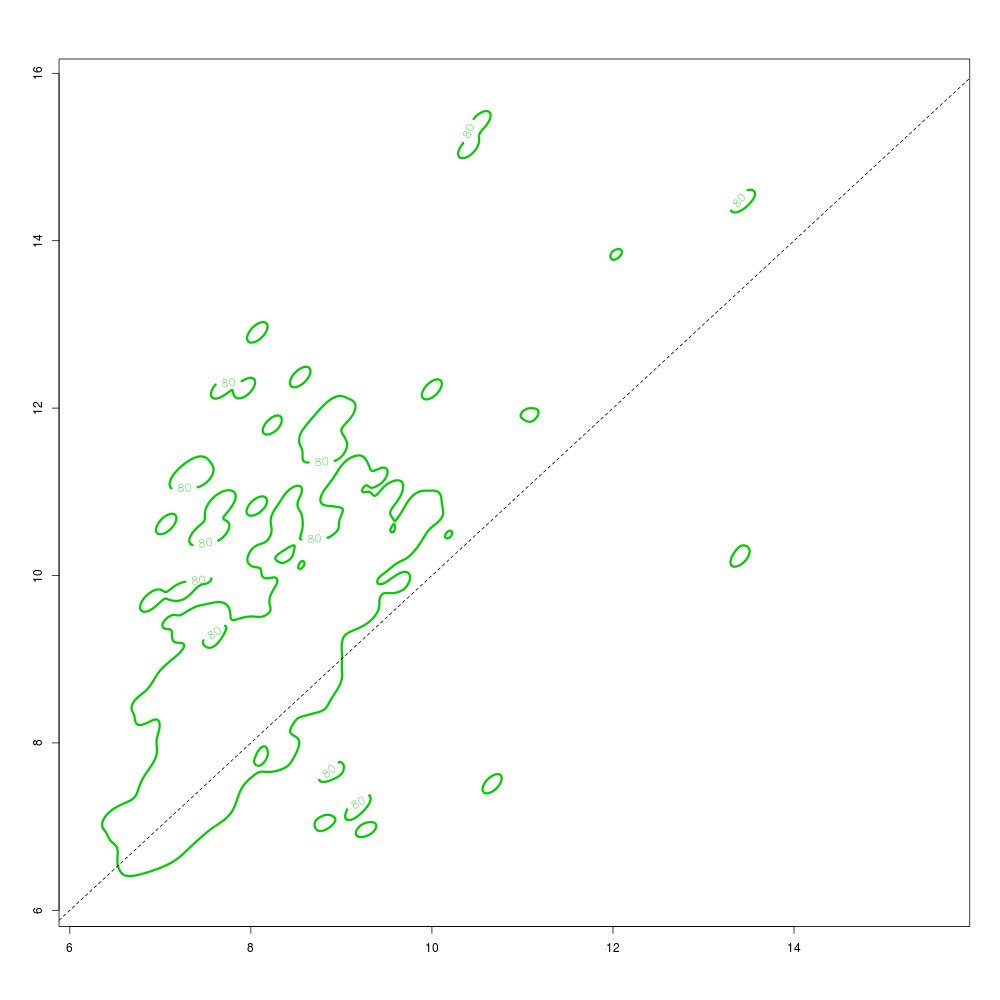} &
    \includegraphics[width=.3\textwidth]{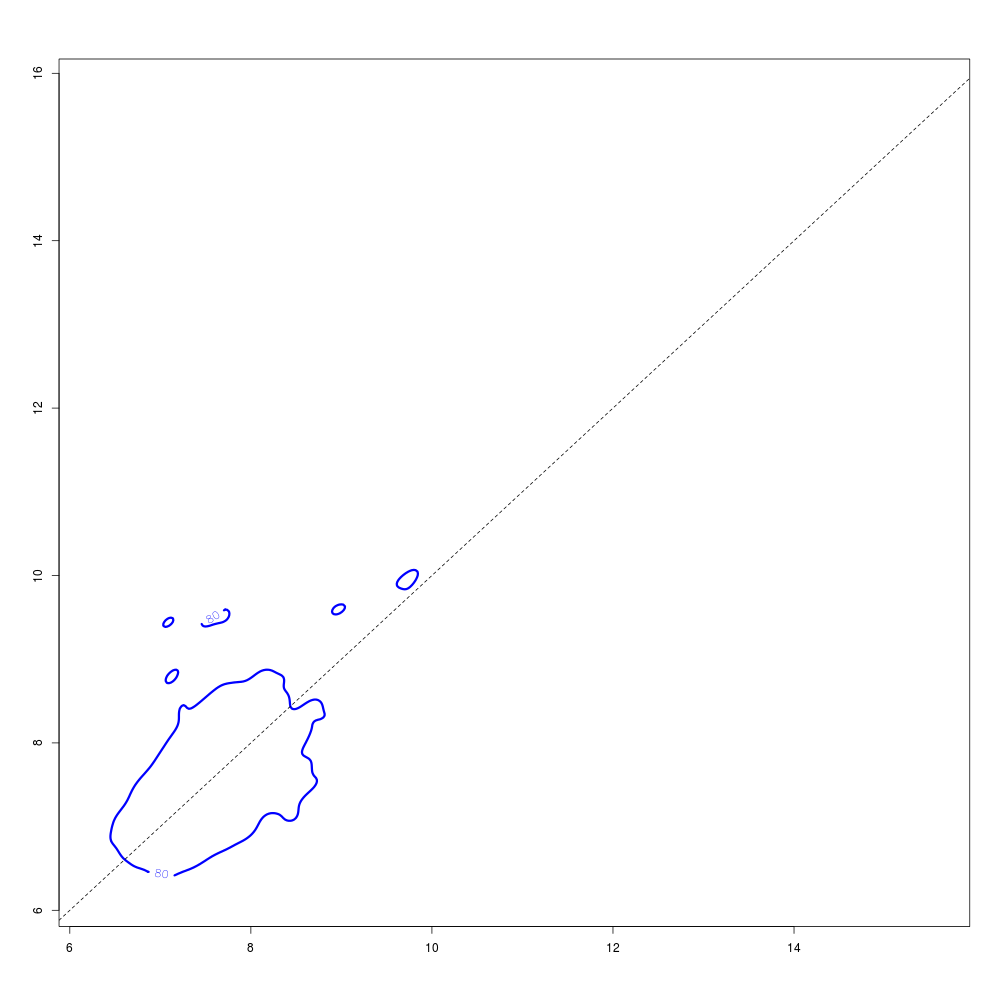} &
    \includegraphics[width=.3\textwidth]{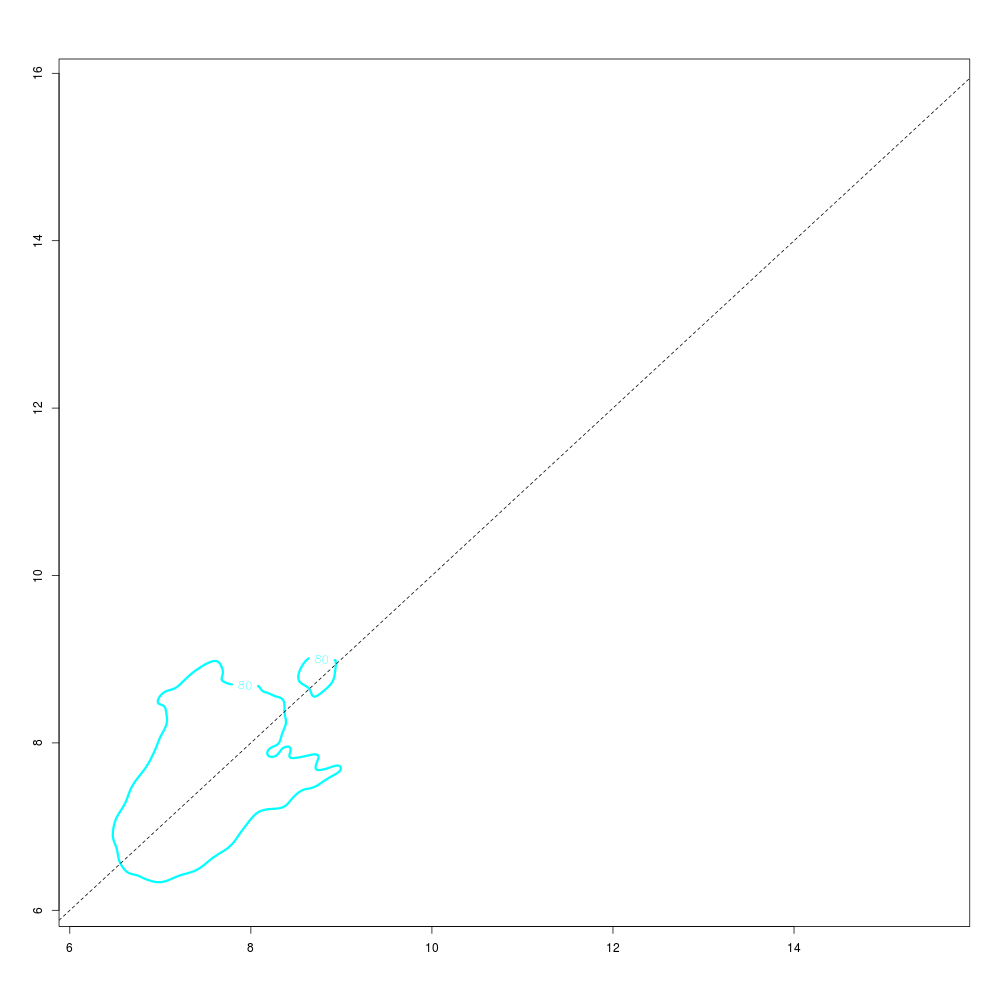} \\
    \end{tabular}
    \caption{Top left panel: Raw tiling array data from chromosome 4 + idealized groups. Other panels: 80\% contour plots of the kernel estimate of each emission distribution for the 5-state non-parametric HMM. The idealized blue group is split into two HMM states (blue and cyan).
    \label{fig:transcriptome}}
  \end{center}
\end{figure}

\paragraph{Tiling array.}
Tiling arrays are a specific microarray technology, where the probes are spread regularly along the genome both in coding and non-coding regions. In transcriptomic applications, tiling arrays capture the intensity of the transcriptional activity at each probe location, thus allowing the detection of transcribed regions. We consider here a comparative experiment were two organs (seed and leaf) of the model plant {\sl A. thaliana} are compared. The data under study corresponds to probes located on chromosome 4. The top left panel of Figure \ref{fig:transcriptome} is an idealization of the expected result. Indeed, we expect to find probes being expressed in none of the organs (blue region), probes being expressed in both organs with equal level (black region) and probes being more expressed in one organ than the other (red and green regions). 

\paragraph{Mixture as emission distributions.}
The same data where already analysed in \cite{VBM13} and \cite{BMB11}, using two different kinds of mixture as emission distributions. The former proposed a very problem oriented mixture of elliptic Gaussian distribution, whereas the latter was a generalization of the approach of \cite{baudry2010combining} to hidden Markov models. A consequence of Proposition \ref{prop:mix} given above is that both of these models are identifiable.

\paragraph{Non-parametric HMM.}
Here, we fitted a $k$-state non-parametric HMM to these data using the kernel method described in Section \ref{subsec:ker}. We used a spherical Gaussian kernal for which we first estimated the bandwidth $w$ via cross-validation on the whole dataset. The 4-state model did not recover the expected structure whereas the 5-state model did, splitting the blue group into two. The rest of Figure \ref{fig:transcriptome} provides the kernel density estimates of the emission distributions under this model. The shape of these distribution turn out to be far from what could be captured by some standard parametric distribution (e.g. 2-dimensional Gaussian). 

\section{Conclusion}
\label{sec:conclusion}

In this article, we have showed that non-parametric hidden Markov models are identifiable up to state-label switching provided that the transition matrix has full rank and that the emission distributions are linearly independent. This gives support to numerous methods that had previously been proposed for the classification of data using non-parametric HMMs. While they usually proved excellent empirical results, no guarantees on the identifiability of the models had yet been given. We provide numerous examples of procedures for which our result applies, and illustrate the gain provided by the use of a non-parametric emission distribution on two examples. In the first example, we present a simulation study inspired from RNA-Seq experiments. In this context, the addition of a regularization function improves the performances of the non-parametric HMM classification. 
In the second example, we present the application of kernel based estimation of emission densities to apply on transcriptomic tiling array data. Again, non parametric estimation improves the classification performances.
This motivates future work on the choices that are involved in non parametric procedures: selection of the regularizing sequence $\lambda_n$ in regularized maximum likelihood,  proposition of a penalty function for the choice of the number of states, choice of mixture components modeling for the emission distribution, choice of the kernel $R$ in a kernel based maximum likelihood estimation, choice of the bandwidth $w$.
\\

{\bf Acknowledgments:} the authors want to thank Caroline B\'erard for providing the transcriptomic tiling array data. 
%\bibliographystyle{../../../../LATEX/astats}
%\bibliography{bibliographie1}

\bibliographystyle{plain}
\bibliography{CGR}

\appendix
\section{Proofs} \label{sec:proofs}

{\bf Proof of Proposition \ref{prop:puj}.}
First remark that $P^{\ell+1} = (p^{\ell+1}_{uj})$ satisfies
$$
P^{\ell+1} = \arg\max_P \sum_{i, u, j} \tau^h_{ij} \gamma^{\ell}_{iuj} \log \left(p_{uj} R_{iu} \right), 
\qquad \text{s.t. }
\forall j: \sum_u p_{uj} = 1.
$$
It follows that
\begin{eqnarray*}
 0 & \leq & \sum_{i, u, j} \tau^h_{ij} \gamma^{\ell}_{iuj} \log \left(p_{uj}^{\ell+1} R_{iu} \right) - \sum_{i, u, j} \tau^h_{ij} \gamma^{\ell}_{iuj} \log \left(p^{\ell}_{uj} R_{iu} \right) \\
 & = & \sum_{i, u, j} \tau^h_{ij} \gamma^{\ell}_{iuj}  \log \frac{p_{uj}^{\ell+1} R_{iu}}{p^{\ell}_{uj} R_{iu}} \quad \leq \quad\sum_{i, j} \tau^h_{ij} \log \left(\sum_u \gamma^{\ell}_{iuj} \frac{p_{uj}^{\ell+1} R_{iu}}{p^{\ell}_{uj} R_{iu}} \right) \qquad \text{(by Jensen's inequality)} \\
% & = & \sum_{i, j} \tau^h_{ij} \log \left(\sum_u \frac1{\sum_v p^\ell_{vj} k_{iv}} p_{uj}^{\ell+1} k_{iu} \right) \\
 & = & \sum_{i, j} \tau^h_{ij} \log \frac{\sum_u p_{uj}^{\ell+1} R_{iu}}{\sum_v p^\ell_{vj} R_{iv}} \quad = \quad G^h(P^{\ell+1}) - G^h(P^{\ell}).
\end{eqnarray*}
$\square$

\end{document}